\documentclass[showpacs,amsmath,amssymb,aps,12pt]{revtex4}
\usepackage{graphicx}
\usepackage{dcolumn}
\usepackage{bm}
\usepackage[latin1]{inputenc}

\begin{document}
\preprint{Preprint/PRB}
\title{Sample Dependence of Structural, Vibrational
and Electronic Properties of $a$-Si:H : A Density
Functional Based Tight Binding Study}

\author{Ranber Singh}
\affiliation{Department of Physics, Panjab University, Chandigarh,
INDIA-160014.}

\author {S. Prakash}
\affiliation{Jiwaji University, Gwalior, INDIA-474011.}

\author{Nitya Nath Shukla and R. Prasad}
\affiliation{Department of Physics, Indian Institute of Technology
Kanpur, INDIA-208016}

\date{\today}
\begin{abstract}
In order to investigate the sample dependence of various
properties of hydrogenated amorphous silicon ($a$-Si:H), we have
generated four samples with 216 silicon atoms and 24 hydrogen
atoms using the density functional based tight binding molecular
dynamics simulations. The overall structural properties of these
model samples are in agreement with the previous theoretical and
experimental results. While the Si-Si and Si-H pair correlation
functions are independent of preparation procedure as well as
initial conditions, the H-H pair correlation functions are sample
dependent. The distribution of hydrogen atoms in all the samples
is nonuniform and depends upon the preparation procedure as well
as the initial structure from which the hydrogenated amorphous
silicon sample is generated. The Si-Si bond length and Si-Si-Si
bond angle distributions are nearly independent of sample
preparation procedure, but Si-H bond length distributions are
sample dependent. The peaks in the vibrational density of states
(VDOS) at high frequencies, which are due to the Si-H bond
vibrations, are in reasonable accord with the experimental
results. The positions of the high frequency peaks are found to be
dependent on the local environment which changes from one sample
to another. While the high frequency vibrational modes related to
Si-Si bond vibrations are moderately localized, the vibrational
modes related to Si-H bond vibrations in $a$-Si:H samples are
highly localized. In samples generated from the liquid quench the
free energy is smaller, whereas the entropy and specific heat are
larger as compared to that of samples generated by hydrogenation
of pure amorphous silicon samples. The total electronic density of
states and local density of electronic states (LDOS) at Si atom
sites are nearly sample independent, while the LDOS at the H atom
sites are dependent on the sample preparation procedure.
\end{abstract}

 \maketitle

\section{Introduction}
Hydrogenated amorphous silicon ($a$-Si:H), technologically an
important material which suffers from the light-induced
degradation [Staebler-Wronski effect \cite{sta77}], has been
studied extensively both experimentally and theoretically
\cite{phy95}. However, the issues related to its various
microscopic properties such as coordination defects \cite{pjp02},
vibrational localization \cite{rel99}, lifetime and decay of
phonons \cite{mvo02}, electronic gap states \cite{for99,nad02},
distribution and dynamics of hydrogen
\cite{raj96,wu96,acc98,will03,tou99,fed08,su02,bis02} are still
the subjects of debate. Several studies show that this material
exhibits structural inhomogeneities \cite{will03,agr96,sin00}. The
study of pure and hydrogenated amorphous silicon is also important
from other perspectives as these are prototypes of disordered
covalent semiconducting materials.
\par
$a$-Si:H can be grown by various techniques including sputtering
\cite{pri02} and plasma enhanced chemical vapor deposition (PECVD)
\cite{pla95}. For experimental studies and practical applications
it is grown in thin films \cite{amor00}, whereas for theoretical
studies it is generated by using systematic computer simulation
techniques. These techniques make use of empirical
\cite{sti85,bis85,ter86} or \emph{ab-intio} pseudo potentials
\cite{bac82} in conjunction with Monte Carlo \cite{wob85,bar00}
and MD simulations \cite{car85,bis87,lue89,qim94} which allow
simulations of thermal treatments similar to those used in
laboratories to prepare the actual samples. Currently, there are
many realistic atomic models of pure amorphous silicon available
(upto 100000 atoms \cite{vin01})  which have been generated by
using the continuous random network forming computer algorithms
\cite{wob85,bar00}. However, due to difficulties in modelling the
interactions of hydrogen in the bulk silicon and a small time step
needed for the dynamics of hydrogen, the modelling of hydrogenated
amorphous silicon is limited and has been achieved mainly by MD
simulations \cite{bud91,zaf91,mou91,fed93,bla96}. Moreover, the
theoretical understanding of its various structural, vibrational
and electronic properties is still not complete and motivates for
extensive studies.
\par
Due to its non-equilibrated nature, $a$-Si:H shows a variation in
its properties depending on the sample preparation conditions. In
this paper we address the question of dependence of various
properties of $a$-Si:H on the preparation conditions. We have
generated four samples of $a$-Si:H with 216 Si atoms and 24 H
atoms using the density functional based tight binding (DFTB)
approach \cite{els00}. Two samples of $a$-Si:H are generated from
the liquid quench at two different quenching rates while the other
two are generated by the hydrogenation of two different pure
amorphous silicon samples. One sample of pure amorphous silicon is
generated from the liquid quench by the same DFTB approach while
the other one is due to Wooten \emph{et al.} \cite{wob85}.
Frauenheim et al. have successfully used the DFTB approach \cite{els00}
to generate the structure of $a$-Si:H (120 Si and 8 H atoms) and
studied its structural and dynamical properties \cite{pet99}. But
in this paper, we studied the sample dependence of the structural, 
vibrational and electronic
properties of various samples of $a$-Si:H generated by using the same DFTB 
approach.
\par
The paper is organized as follows. 
The computational details of methods used for
generating samples and subsequent calculations of dynamical
matrix, localization of vibrational modes, VDOS, thermodynamic
properties from the VDOS and electronic density of states are
given in Sec II. The results of our calculations are discussed in
Sec III and summarized in Sec IV.

\section{Computational Details}
We have used DFTB molecular dynamics scheme of Frauenheim \emph
{et al.} \cite{els00} to generate the structures of pure and
hydrogenated amorphous silicon and to calculate their vibrational
and electronic properties.
For the structural relaxation, we sample the Brillouin zone of the
supercell lattice at the $\Gamma $ point alone and adopt the
canonical ensemble scheme where the kinetic energy of the ionic
motions is constantly rescaled to the simulation temperature. In
the stimulated annealing mode of the MD program, the force
tolerance threshold is set to 0.0001 (atomic units) and the
electronic temperature for Fermi broadening to 300 K. We used
sp-basis set for Si and s-basis set for H. The range of repulsive
potential and cut-off distance for the next nearest neighbour (in
atomic units) respectively are 4.8 and 4.8 for Si-Si interactions,
3.2 and 3.2 for Si-H interactions, 2.99 and 2.0 for H-H
interactions. The ionic motions are calculated with discrete time
steps of about 1 fs. For the generation of structures of pure and
hydrogenated amorphous silicon from liquid quench, we used the
dynamical method. In this method we start with crystalline sample
then raise the temperature of the sample to 4000 K. The sample is
then equilibrated at 4000 K for 0.5 ps to destroy the memories of
its initial structure and then cooled down to 1800 K in 2 ps to
form a liquid where it is equilibrated for further 2 ps. The cell
is then quenched to 300 K where it is equilibrated for some more
time. The samples of pure and hydrogenated amorphous silicon with
216 silicon atoms and 24 hydrogen atoms are generated within a
simple cubic simulation cell of side 16.2813 {\AA}. We have taken
the simulation cell of size 16.2813 {\AA} to have the same mass
density of pure amorphous silicon sample as that of WWW model
sample \cite{wob85}. The hydrogen atoms are added into the initial
crystalline silicon by dividing the simulation cell into 27 equal
cubes and placing hydrogen atoms at tetrahedral positions, one in
each of 24 of the 26 outer cubes. While two samples of $a$-Si:H
are generated from the liquid quench at two different quenching
rates, two more samples are generated
by the hydrogenation of pure amorphous silicon samples. These samples are 
labelled as follows:\\
$\bf{ASi}$: Pure amorphous silicon sample generated by using a
quenching rate of $2\times10^{14}$
K/s from the liquid state at 1800 K to 300 K where it is further 
equilibrated for 2.5 ps.\\
$\bf{ASiH}$: $a$-Si:H sample generated by using a quenching rate
of $2\times10^{14}$ K/s from the liquid state at 1800 K to 300 K
where it is  further equilibrated for 2.5 ps. At 300 K the
diffusion almost ceases. Therefore, to relax the structure further
the temperature is raised to 1200 K, equilibrated there for 1 ps
and then cooled again to 300 K at
the rate of $4\times10^{14}$ K/s. At 300 K, it is equilibrated for 2.0 ps.\\
$\bf{BSiH}$: $a$-Si:H sample generated by using a quenching rate
of $4\times10^{14}$ K/s from the liquid state at 1800 K to 1200 K
where it is equilibrated for 1 ps and then cooled down to 300 K at
the same rate. At 300 K it is further equilibrated for 5 ps. The
structure is further relaxed by raising the temperature to 1200 K
where it is equilibrated for 1 ps and then cooled to 300 K at the
rate of $4\times10^{14}$ K/s. At 300 K, it is again equilibrated
for 2.0 ps.\\
$\bf{CSiH}$: $a$-Si:H sample generated by hydrogenation of {\bf
ASi} sample of pure amorphous silicon. This is done by dividing
the simulation cell into 27 equal cubes and placing hydrogen atoms
at arbitrary positions in 24 of the 26 outer cubes. The resultant
structure is simulated for 0.5 ps at 300 K. To increase the
diffusion of atoms, the temperature is raised to 1200 K at the
rate of $6\times10^{14}$ K/s where it is equilibrated again for 1
ps and then cooled to 300 K at the rate of $2\times10^{14}$ K/s.
At 300 K, it is again equilibrated for 2.5 ps. \\
$\bf{WSiH}$: $a$-Si:H sample generated by hydrogenation of WWW
sample of pure amorphous silicon of Wooten \emph{et al.}
\cite{wob85} adopting the same procedure as for CSiH.
\par
To investigate the vibrational properties of the above samples,
the dynamical matrix is constructed with the help of the
differentiation scheme defined in Eq.(9) with $\Delta $=0.005
{\AA}. The diagonalization of this dynamical matrix gives the
eigenvalues (square of vibrational frequencies) and the
eigenvectors (vibrational patterns) which are further used to
calculate the vibrational density of states (VDOS) and
localization of vibrational modes of the model samples. The
broadening width of 20 cm$^{-1}$ is used to calculate the VDOS.
Once the information of VDOS is available, it is easy to construct
the temperature dependence of thermodynamic properties. In the
harmonic approximation the free energy (F), entropy (S) and
specific heat (C) per atom of a simulated system are calculated
using expressions given in Ref. \cite{mar71}.
\par
The localization behaviour of the vibrational modes is
investigated by calculating the inverse participation ratio (IPR)
as given in Ref. \cite{bis88}.
A vibrational mode for which the participation
ratio is approaching 1 will be a localized mode.
For the electronic structure calculations of the above model
structures, a huge memory is required. Therefore, we could use the
maximum of 16 k-points of the Brillouin zone. For the calculation
of electronic density of states we have used $2\times2\times2$
Monkhorst-Pack \cite{mon76} mesh of k-points.
\section{Results and Discussion}
\subsection{Structural Properties}
We first discuss the structural properties of the four $a$-Si:H
samples with the help of Figs.1-5 and Tables I-III. The data, as
given in these figures and tables, are the averages of 100
configurations over a time span of 0.5 ps. The pair correlation
function for the pure amorphous silicon as given in Fig.1 shows a
good agreement with the experimental data \cite{laz99}. The first
peak which corresponds to the first nearest neighbours is sharp
and occurs at 2.37 {\AA}. However, the second and third peaks
corresponding to the second and third nearest neighbours are
broader. The partial pair correlation functions of Si-Si, Si-H and
H-H correlations in $a$-Si:H samples (ASiH, BSiH, CSiH and WSiH)
are given in Fig.2. In Si-Si and Si-H pair correlation functions,
the first peaks corresponding to Si-Si and Si-H bond lengths are
sharp. However, in H-H pair correlation function it is broadened
and occurs above 2.0 {\AA} which indicates the nonexistence of
molecular hydrogen (bond length $\approx$ 1.0 {\AA}) but the
hydrogen configurations with two hydrogen atoms spaced more than
2.0 {\AA} apart and/ or clustering of hydrogen atoms in $a$-Si:H.
This is in agreement with the \emph{ab initio} results
\cite{bud91} as well as the experimental data \cite{bel89}. The
second and third nearest neighbour peaks in Si-Si and Si-H
correlation functions in the $a$-Si:H sample are broader. In H-H
correlation functions there is no clear peak except the first
peak. The positions of main peaks in the pair correlation
functions of different samples are given in Table I, where the
data from the previous theoretical \cite{wob85,bud91,gup97} and
experimental \cite{laz99,bel89} studies are also given for
comparison. There is a good agreement between the positions of
peaks in the pair correlation functions of our samples and the
experimental data.
\par
From Fig.2, it is also clear that as far as Si-Si and Si-H pair
correlation functions are concerned all the four samples (ASiH,
BSiH, CSiH and WSiH) are quite similar. This shows that Si-Si and
Si-H pair correlation functions are independent of the preparation
procedure and initial structure from which the hydrogenated sample
is generated. However, the H-H pair correlation functions show the
dependence on preparation procedure. The overall
behaviour of H-H pair correlation functions is quite similar to
the experimental results \cite{bel89}. However, due to the finite
size of the samples and a small number of hydrogen atoms, the
statistics is not good enough to draw any general conclusion
except that the distribution of hydrogen in amorphous silicon is
nonuniform and depends upon the preparation procedure and initial
structure from which the hydrogenated sample is generated. A small
angle neutron scattering (SANS) study \cite{will03} of nonuniform
hydrogen distribution in $a$-Si:H has also reported the existence
of hydrogen heterogeneity on nanometer scale in $a$-Si:H.
\par
A more detailed structural description of the model samples is
given by the bond length and bond angle distributions. These
distributions for pure and hydrogenated amorphous silicon samples
are calculated by defining the cut-off distances for Si-Si and
Si-H bondings upto the first minima in their respective
correlation functions. The Si-Si bond length and Si-Si-Si bond
angle distributions for ASi and WWW samples are shown in Fig 3.
These distribution functions show a broad distribution with peaks
respectively at around 2.37 {\AA} and 106$^{o}$. The distribution
functions for ASi and WWW samples are in good agreement except
some noise in the curves corresponding to WWW sample. This is
because of the fact that for WWW sample only one configuration was
used while for ASi sample the average was taken over 100
configurations. These distributions indicate that ASi sample
deviates from the perfect tetrahedral network and has a
distribution over bond length and bond angle. The root mean square
deviations (RMD) of Si-Si bond length and Si-Si-Si bond angle
distributions in ASi and WWW samples are 0.13 {\AA} and
16.65$^{o}$, and 0.08 {\AA} and 11.19$^{o}$ respectively.
\par
Fig.4 shows that the Si-Si bond length and Si-Si-Si bond angle
distribution functions in WSiH sample have sharper peaks as
compared to other samples. These peaks occur respectively at about
2.34 {\AA} and 108$^{o}$ which are quite close to the
corresponding values in crystalline silicon. It indicates that the
WSiH sample has crystalline features which is due to the already
present crystalline components in the model structure of WWW
sample as has been shown by Wooten \emph{et al.} \cite{wob85}.
These distribution functions in ASiH, BSiH and CSiH samples are
nearly identical with a little variation in their peak heights and
positions occurring at around 2.36 {\AA} and 106$^{o}$
respectively. Thus, Si-Si bond length and Si-Si-Si bond angle
distributions are almost independent of preparation procedure and
the initial structure from which the hydrogenated sample is
generated. The RMD of Si-Si bond length and Si-Si-Si bond angle
distributions in ASiH, BSiH, CSiH and WSiH samples are 0.12 {\AA}
and 17.15$^{o}$, 0.11 {\AA} and 17.37$^{o}$, 0.13 {\AA} and
15.79$^{o}$, and 0.11 {\AA} and 15.61$^{o}$ respectively.
\par
The Si-H bond length distributions (see Fig.5) are different in
all the samples except that the peak occurs nearly at about 1.52
{\AA}. In $a$-Si:H samples generated from the liquid quench, this
distribution is broader whereas it is sharper in the samples
prepared by hydrogenation of pure amorphous silicon. This can be
attributed to the stability of Si-H bonding in these samples. As
these distribution functions are the time averages, the broader
distribution indicates that Si-H bonding is unstable so that bond
length has a variation over a certain range, while the narrow
distribution indicates that the Si-H bonding is relatively stable
and there is little variation in the bond length. The comparison
of the Si-H bonding distributions for ASiH and BSiH samples also
shows that the distribution is sharper for BSiH sample prepared by
the faster quenching rate. The Si-Si-H bond angle distributions
(see Fig.5) are nearly similar in all samples except that it is
sharper in WSiH sample. The sharper peak is due to the presence of
crystalline components in the structure of WSiH sample as shown
above by the Si-Si bond length and Si-Si-Si bond angle
distribution functions given in Fig.4. Thus, Fig.5 indicates that
while the Si-H bond length distributions are dependent upon the
preparation procedure as well as the initial structure from which
the hydrogenated sample is generated, the Si-Si-H bond angle
distributions are nearly independent.
\par
In addition to the 4-fold Si atoms there are also 3-fold and
5-fold Si atoms in our samples as given in Table II. In ASi sample
the average percentages of 3-fold, 4-fold and 5-fold Si atoms are
respectively 3.08, 90.85 and 5.91, while in WWW model these are
0.00, 99.10 and 0.90. Upon hydrogenation, these values become
nearly similar in both samples, i.e., 2.41, 95.52 and 2.07 in CSiH
sample and 1.84, 95.85 and 2.29 in WSiH sample. The $a$-Si:H
samples (ASiH, BSiH) generated from the liquid quench have more
coordination defects (3-fold and 5-fold atoms) than those (CSiH,
WSiH) generated by hydrogenation of pure amorphous silicon
samples. Thus the samples generated from the liquid quench are
found to be more disordered as compared to those generated by
hydrogenation of pure amorphous silicon samples. The comparison of
coordination defects in ASiH and BSiH samples shows that the
sample (BSiH) prepared by the faster quenching rate has less
coordination defects than that prepared at the slower rate.
\par
The mean coordination numbers, Z$_{Si-Si}$, Z$_{Si-H}$,
Z$_{H-Si}$, and Z$_{tot}$(=Z$_{Si-Si}$+Z$_{Si-H}$) are given in
Table II. The data available from the previous theoretical studies
\cite{wob85,bud91} are also given for comparison. The total
coordination number, Z$_{tot}$ of a Si atom in all the samples is
nearly equal to 4. The Z$_{Si-Si}$ of a Si atom in ASi sample is
4.03, which is quite close to 4.01 of WWW model \cite{wob85}.
However, upon hydrogenation it reduces to 3.89 in both the samples
(CSiH and WSiH). The Z$_{Si-Si}$ and Z$_{tot}$ of a Si atom in
ASiH and BSiH samples is larger than that of CSiH and WSiH
samples. This indicates that a Si atom in ASiH and BSiH samples is
more over-coordinated as compared to that in CSiH and WSiH
samples. The Z$_{H-Si}$ of a hydrogen atom in all the $a$-Si:H
samples is nearly equal to 1.0 which means that almost all the
hydrogen atoms are bonded to silicon atoms as Si-H monohydrides.
This is in agreement with the vibrational spectroscopy results
\cite{bro77,luc79}. Further, we found that Si-H monohydrides exist
either as isolated Si-H configurations or as the clusters of Si-H
monohydrides. The Si-H clustering has also been observed in
experiment \cite{bjd00} and computer simulations
\cite{bud91,bla96,acc98,sme03}.
\par
The average values of mean bond lengths (Si-Si \& Si-H) and mean
bond angles (Si-Si-Si, Si-Si-H, Si-H-Si \& H-Si-H) are given in
Table III. The Si-Si bond length in ASi sample is 2.38 {\AA}
compared to 2.35 {\AA} in WWW model and 2.35 {\AA} in crystalline
silicon. In the $a$-Si:H samples viz ASiH, BSiH, CSiH and WSiH, it
is respectively 2.39, 2.38, 2.36 and 2.36 {\AA}. This shows that
in ASiH and BSiH samples, the Si-Si bond length is nearly equal to
that in the ASi sample, but in other hydrogenated samples it is
smaller. This decrease in  Si-Si bond length can be attributed to
the relaxation of strain associated with the weak Si-Si bonds upon
hydrogenation. The Si-H bond length in ASiH, BSiH, CSiH and WSiH
samples is 1.56, 1.54, 1.53 and 1.53 {\AA} respectively. The
comparison of Si-Si and Si-H bond lengths in $a$-Si:H samples
indicates that these bond lengths are larger in the samples (ASiH,
BSiH) prepared from the liquid quench compared to the samples
(CSiH, WSiH) prepared by hydrogenation of pure amorphous silicon
samples. This means that the samples prepared by hydrogenation of
pure amorphous silicon have less strain compared to the samples
prepared from liquid quench. However, the comparison of Si-Si and
Si-H bond lengths in ASiH and BSiH samples indicates that BSiH
sample has less strain compared to ASiH sample. The mean Si-Si-Si
bond angle in ASi sample is 108.23$^{o}$ compared to 109.12$^{o}$
in WWW model and 109.47$^{o}$ in crystalline silicon. In ASiH and
BSiH samples it is nearly equal to 108$^{o}$, while in CSiH and
WSiH samples it is 108.8$^{o}$. This indicates that CSiH and WSiH
samples are closer to the perfect tetrahedral network. The mean
Si-Si-H bond angle is less than Si-Si-Si bond angle, i.e., 103.20,
103.81, 104.67 and 102.79$^{o}$ respectively in ASiH, BSiH, CSiH
and WSiH samples. This means that three Si atoms bonded to a
hydrogen bonded Si atom have the tendency to have a planar
structure. There exists stable Si-H-Si bridge configuration in
ASiH and WSiH samples with Si-H-Si bond angle nearly equal to
157$^{o}$. Such stable Si-H-Si bridge configuration does not exist
in BSiH and CSiH samples. However in CSiH sample, it forms, stays
for some time and then breaks with the average Si-H-Si angle of
114.36$^{o}$ over a time span of 0.5 ps. In ASiH and BSiH samples,
there also exist SiH$_{2}$ configurations with average H-Si-H bond
angle equal to 107.90$^{o}$ and 102.25$^{o}$ respectively.
\subsection{Vibrational Properties}
The VDOS of ASi sample of pure amorphous silicon is given in
Fig.6(a). Also shown in Fig.6.(a) are experimental \cite{kam87} and
theoretical \cite{nak98,nak00} results. Our calculations are in
good agreement with the experimental data in the whole frequency
range. We note that previous theoretical results
\cite{nak98,nak00} do not compare well with the experimental
results. In the fireball-DTW \cite{nak98} results, the low
frequency peak is larger while in edip-DTW \cite{nak00} results it
is smaller than that of the experimental data. However, the high
frequency peak in VDOS of both fireball-DTW and edip-DTW results
is at higher frequencies compared to the experimental data. The
IPR of vibrational modes (see Fig.6(b)) shows that the high
frequency modes are localized in ASi sample. Similar conclusions
have also been reported by other authors
\cite{bis88,nak98,pbis02}. Due to the finite size of our ASi
sample, the vibrational modes below 50 cm$^{-1}$ are missing but
we have found one localized mode at around 30 cm$^{-1}$. Such
localized mode could arise due to the presence of voids in the
sample as has been reported by Nakhmanson \emph{et al.}
\cite{nak98} in amorphous silicon.
\par
The calculated VDOS and IPR of vibrational modes in the
hydrogenated samples (ASiH, BSiH, CSiH and WSiH) are shown in
Fig.7. For clarity, we have displayed them separately for each
sample. The main peaks in VDOS of all the hydrogenated samples lie
below 550 cm$^{-1}$ and are almost similar. The small peaks due to
the Si-H vibrational modes above the main peaks lie in the
frequency range of 600-900 cm$^{-1}$ and 1900-2200 cm$^{-1}$ in
all samples and their locations change from sample to sample.
There are three types of Si-H vibrational modes in amorphous
silicon: (i) wagging modes (630-650 cm$^{-1}$), (ii) bending modes
(830-860 cm$^{-1}$), and (iii) stretching modes (2000-2200
cm$^{-1}$). The dominating peaks are at around 626, 840, 2053 and 2107 
cm$^{-1}$ in ASiH;
661, 830, 2035, 2085 cm$^{-1}$ in BSiH; 645, 2054, 2117 cm$^{-1}$
in CSiH and 646, 2049, 2114 cm$^{-1}$ in WSiH. This indicates that
the Si-H vibrations are sensitive to the local environment as it
changes from one sample to another. The experimental measurement
of VDOS of $a$-Si:H \cite{kam84} has shown the existence of extra
peaks in VDOS at about 646, 889, 2000, 2100 cm$^{-1}$ respectively
which have been attributed to the wagging modes of hydrogen,
bending modes of SiH$_{2}$ configuration, stretching mode of SiH
and stretching mode of SiH$_{2}$ vibrations. The Raman spectra of
$a$-Si:H \cite{mil91} shows these peaks at 640, 2000, 2100
cm$^{-1}$. Mauseau \emph{et al.} \cite{nor90} also calculated the
VDOS of hydrogen atoms in $a$-Si:H and identified three main peaks
at 645, 888, 2100 cm$^{-1}$. In the IR spectra of $a$-Si:H
\cite{tou99,bro77,lko96}, the (Si-H)$_{x}$ hydride vibrations have
characteristic absorptions at 630-650 cm$^{-1}$ (wagging mode),
840-890 cm$^{-1}$ (bending scissor mode of H-Si-H), 1980-2030
cm$^{-1}$ (Si-H stretching mode), 2060-2160 cm$^{-1}$ (H-Si-H
stretching mode and /or clusters of Si-H monohydrides).
\par
The IPR of vibrational modes in different hydrogenated samples in
Fig.7 shows that the localization of vibrational modes below 575
cm$^{-1}$ in all the hydrogenated samples is almost similar to
that in the pure amorphous silicon sample (Fig.6(b)). However, the
vibrational modes above 575 cm$^{-1}$ which are mainly related to
Si-H bond vibrations are highly localized. Recently, Rella
\emph{et al.} \cite{rel99} have also reported that Si-H stretch
vibrations are highly localized. The comparison of IPR of Si-H
vibrational modes in different samples also indicates that their
localization is sensitive to the local environment as it changes
from one sample to another. Similar to ASi sample, in $a$-Si:H
samples also there are some localized modes below 50 cm$^{-1}$
which may be arising due to the presence of voids in the samples.
\par
In the amorphous solids there exist some universal features, for
example, a bump in the low-temperature specific heat \cite{sok94}
and an excess of low frequency modes (about 20-120 cm$^{-1}$) in
the Raman and neutron scattering spectra \cite{mmsf01}. In the
neutron scattering spectra, it is usually expressed as a peak in
VDOS/$\omega^{2}$. We displayed VDOS/$\omega^{2}$ plots for our
samples of pure and hydrogenated amorphous silicon in Fig.8. We
find that there occurs a broad band peak with two small bumps in
VDOS/$\omega^{2}$ plots near 100 cm$^{-1}$ in all the samples. In
ASi sample, there is a broad bump at around 95 cm$^{-1}$ while in
$a$-Si:H samples viz: ASiH, BSiH, CSiH and WSiH, it is around 107,
104, 105 and 105 cm$^{-1}$ respectively. The Raman scattering on
bulk silicon \cite{lae99} has also reported the peak at 114
cm$^{-1}$. We notice that the position of peak is almost
independent of the sample and depends only on the type of sample
whether pure or hydrogenated amorphous silicon. In $a$-Si:H
samples, the height of peak is smaller as compared to that in pure
amorphous silicon sample (ASi) which means that there is a
decrease in the low frequency modes in amorphous silicon upon
hydrogenation. Liu \emph{et al.} \cite{xia97} have also reported
such a reduction in the low frequency modes in amorphous silicon
upon hydrogenation.
\par
The VDOS is further used to calculate the temperature (T)
dependence of free energy, vibrational entropy and specific heat
(C). The overall temperature
dependence of specific heat for our samples is in good agreement
with the Dulong and Petit's law at high temperatures. From Fig.9,
it is clear that while the free energy decreases, the entropy and
specific heat increase with the increase in temperature in all the
samples. In the $a$-Si:H samples, the free energy is more, whereas
the entropy and specific heat are less than those of the pure
amorphous silicon sample at all temperatures. However, the
inter-comparison of $a$-Si:H samples shows that in samples (ASiH,
BSiH) prepared from the liquid quench, the free energy is smaller
whereas the entropy and specific heat are larger as compared to
that of samples (CSiH, WSiH) prepared by hydrogenation of pure
amorphous silicon samples at all the temperatures. Since the
entropy is the measure of disorder, the higher entropy of pure
amorphous sample (ASi) indicates that it is more disordered
compared to the hydrogenated samples. The comparison of entropy of
hydrogenated samples shows that the samples generated from liquid
quench are more disordered compared to other samples.
\par
In Fig.9(d), we have displayed C/T$^{3}$ versus T curves. The
inset in Fig.9(d) shows a clear bulge in these curves at low
temperatures in all of our samples. The existence and position of
this bulge in the specific-heat at low temperatures in our samples
is in good agreement with the previous theoretical
\cite{fel99,snak00} and experimental results \cite{mer84}. The
height of the peak in the hydrogenated samples is lower as
compared to that in the pure amorphous silicon sample. The
comparison of heights of peak in the hydrogenated samples shows
that it is lower in the samples (ASiH, BSiH) prepared by
hydrogenation of pure amorphous silicon samples as compared to
that in other samples prepared from the liquid quench. Although
ASiH and BSiH samples have quite different coordination defects,
they have almost identical peak heights in C/T$^{3}$ plots, while
CSiH and WSiH samples have almost similar coordination defects but
they have quite different peak heights in C/T$^{3}$ plots. This
indicates no connection between coordination defects and the peak
in specific heat at low temperatures.
\subsection{Electronic Properties}
We have investigated the electronic properties of pure and
hydrogenated amorphous silicon samples by calculating the
electronic density of states. The calculated results are given in
Figs.10-11. The energy, E=0 eV in all the panels of these figures
corresponds to the Fermi level of the corresponding samples as
shown in the panels. The electronic density of states shows a
small energy gap at the Fermi level in all of our samples which is
quite typical for such model structures as has also been found in
various previous theoretical calculations \cite{for99,bpi03}. The
total electronic density of states (EDOS) of pure amorphous
silicon samples (ASi and WWW) as given in Fig.10 shows that there
is a clear formation of a gap around the Fermi level in both the
samples. The general shape of EDOS in both the samples is almost
the same except that the band gap around the Fermi level is wider
in WWW sample as compared to that in ASi sample. Thus there are
more states in the gap region of ASi sample as compared to that in
WWW sample. As is well known, these gap states are due to the
presence of coordination defects. The ASi sample has about 9\%
coordination defects (both 3-fold and 5-fold), while the WWW
sample has only 0.1\% 5-fold defects. The shape of band tails near
the gap region is exponential in both of these samples. Such an
exponential behaviour of band tail states in amorphous silicon has
also been shown by Jianjun \emph{et al.} \cite{jia98} and is a
universal feature of amorphous semiconductors caused by the short
range disorder in these materials.
\par
The local electronic density of states of hydrogen atoms
(LDOS$_{H}$) and Si atoms (LDOS$_{Si}$) and total electronic
density of states (EDOS) for the $a$-Si:H samples are given in
Fig.11. For clarity we displayed them separately for each sample.
In valance band region (-10.0 - 0.0 eV) LDOS$_{H}$ has a similar
shape in all the samples but in the conduction band region (0.0 -
10.0 eV) has a different shape for the samples (ASiH, BSiH)
generated from the liquid quench and for those (CSiH, WSiH)
generated by hydrogenation of pure amorphous silicon samples. The
general shape of LDOS$_{Si}$ at Si sites and EDOS is almost the
same in all of the four samples. The band tail states near band
gap region in EDOS have almost the same exponential distribution
as in the pure amorphous silicon sample. There is a decrease in
the electronic band tail states in the gap region and increase in
the energy gap around Fermi level in EDOS in the hydrogenated
samples which is consistent with the earlier studies \cite{phy95}.
Balamurugan and Prasad \cite{bal01} also reported the increase of
first excited electronic level gap in the small silicon clusters
upon hydrogenation. 
Thus hydrogenation has the
tendency to remove the states from the gap region and increase the
band gap.
\section{Conclusions}
We have presented the DFTB based MD calculations for the sample
dependence of structural, vibrational and electronic
properties $a$-Si:H. The samples of $a$-Si:H are generated by quenching from 
the liquid
state of silicon-hydrogen mixture and by hydrogenation of pure
amorphous silicon samples. The overall structural properties of
these model structures are in good agreement with the previous
experimental and theoretical results. The $a$-Si:H samples
generated from the liquid quench have more coordination defects as
compared to the samples generated by hydrogenation of pure
amorphous silicon. The Si-Si and Si-H pair correlations are found
independent of preparation procedure and initial conditions,
whereas the H-H pair correlations are sample dependent. It has
been found that almost all the hydrogen atoms are bonded to
silicon atoms, in Si-H monohydride fashion, existing either as
isolated Si-H bond or as clusters of Si-H bonds. The distribution
of hydrogen in all the samples is found nonuniform and depends
upon the preparation procedure and the initial structure from
which the hydrogenated sample is generated. The Si-Si bond length
and Si-Si-Si bond angle distributions are nearly independent of
sample preparation procedure, but Si-H bond length distributions
are dependent.
\par
The vibrational properties of these samples are investigated by
calculating the VDOS and localization of vibrational modes. The
VDOS of pure amorphous silicon sample is in good agreement with
the experimental data. In hydrogenated samples, there exist extra
peaks in the frequency range 600 - 2200 cm$^{-1}$ which are also
in reasonable accord with the experimental results. In
VDOS/$\omega^{2}$ versus $\omega$ plot there occurs a peak at low
frequencies in all of our samples. The position and height of this
peak shows no connection to coordination defects. The height of
this peak also indicates that the low frequency vibrational modes
are less in the samples generated by hydrogenation of pure
amorphous silicon samples as compared to the other samples
generated from the liquid quench. While the high frequency modes
related to Si-Si bond vibrations are found moderately localized
both in pure and hydrogenated amorphous silicon, the vibrational
modes related to Si-H vibrations are highly localized.
\par
The VDOS is further used to calculate the temperature dependence
of thermodynamic properties within the harmonic approximation. It
has been found that the free energy decreases whereas the entropy
and specific heat increase with increase in temperature in all
samples. In $a$-Si:H samples, the free energy is larger while the
entropy and specific heat are lower than that of pure amorphous
silicon sample at all the temperatures. However, the
inter-comparison of $a$-Si:H samples shows that in samples
prepared from the liquid quench, the free energy is smaller
whereas the entropy and specific heat are larger as compared to
that of samples prepared by hydrogenation of pure amorphous
silicon samples at all temperatures. At low temperatures, there
exists a bump in C/T$^{3}$ plots of all the samples. The height
and position of this bump show no relation with the coordination
defects. The electronic density of states shows a small energy gap
at the Fermi level in all the samples. The hydrogenation of pure
amorphous silicon reduces the electronic gap states and increases
the energy gap. The total electronic density of states and local
density of electronic states (LDOS) at Si atom sites are nearly
sample independent, while the LDOS at the H atom sites is
dependent on the sample preparation procedure.
\begin{acknowledgments}
We thank Prof. Th. Frauenheim for providing the DFTB code and D.
Balamurugan and Prof. S. C. Agarwal for the fruitful discussions.
We are also thankful to Fred Wooten, R. L. C. Vink, S. M.
Nakhmanson, S. Roorda and W. A. Kamitakahara for sharing their
data. NNS would like to thank Prof. Th. Frauenheim for the
hospitality at Paderborn, Germany during summer 2002 and RS
gratefully acknowledges the financial support from the Council of
Scientific and Industrial Research, Govt. of India, New Delhi.
This work is supported by Department of Science and Technology,
New Delhi under project no. SP/S2/M-51/96.
\end{acknowledgments}

\newpage
\begin{table}[ht]
\caption{\label{tab:table1}The positions of peaks in Si-Si, Si-H
and H-H pair correlation functions of our samples as shown in Figs
1-2. All the values are in {\AA}. The data available from the
previous theoretical \cite{wob85,bud91,gup97} and experimental
\cite{laz99,bel89} studies are also given for comparison.}
\begin{ruledtabular}
\begin{tabular}{c|ccccccccc}
&&Si-Si&&\vline&&Si-H&&\vline&H-H\\
\hline
samples/peaks &I&II&III&\vline&I&II&III&\vline&I\\
\hline
amorphous silicon \cite{laz99}&2.36&3.77&5.74&&&&&&\\
WWW \cite{wob85} &2.36&3.82&5.80&&&&&&\\
ASi&2.37&3.77&5.75&&&&&&\\
$a$-Si:H \cite{bel89} &2.37&3.78&5.64&&1.48&3.13&4.85&&2.38-2.70\\
$a$-Si:H \cite{bud91} &2.35&3.80&-&&1.59&3.12&-&&2.20-2.60\\
$a$-Si:H \cite{gup97} &2.37&3.72&-&&1.59&3.18&5.25&&2.00-4.00\\
ASiH&2.37&3.78&5.71&&1.54&3.10&4.91&&2.31-2.66\\
BSiH&2.36&3.83&5.77&&1.54&3.09&4.97&&2.21-2.38\\
CSiH&2.36&3.81&5.65&&1.53&3.12&4.91&&2.30-2.48\\
WSiH&2.35&3.80&5.79&&1.53&3.04&4.91&&2.20-2.43\\
\end{tabular}
\end{ruledtabular}
\end{table}

\begin{table}[ht]
\caption{\label{tab:table2} The average percentages of n-fold (3,
4, 5) Si atoms and the mean coordination numbers Z$_{Si- Si}$,
Z$_{Si-H}$, Z$_{H-Si}$ and Z$_{tot}$(=Z$_{Si-Si}$+Z$_{Si-H}$) in
our samples. The average is taken of 100 configurations over a
time span of 0.5 ps. The available data from the previous
theoretical \cite{wob85,bud91} studies are also given for
comparison.}
\begin{ruledtabular}
\begin{tabular}{c|cccccccc}
samples&3 & 4 & 5 &\vline&Z$_{Si-Si}$&Z$_{Si-H}$&Z$_{H-Si}$&Z$_{tot}$\\
\hline
WWW \cite{wob85} &0.00&99.10&0.90&&4.01&-&-&4.01\\
ASi&3.08&90.85&5.91&&4.03&-&-&4.03\\
$a$-Si:H \cite{bud91} &1.00&91.50&7.50&&3.90&0.13&1.00&4.03\\
ASiH&4.47&88.20&7.33&&3.92&0.12&1.04&4.04\\
BSiH&3.58&90.81&5.61&&3.91&0.11&1.00&4.02\\
CSiH&2.41&95.52&2.07&&3.89&0.11&1.00&4.00\\
WSiH&1.84&95.85&2.29&&3.89&0.12&1.04&4.01\\
\end{tabular}
\end{ruledtabular}
\end{table}

\begin{table}[ht]
\caption{\label{tab:table3}The average mean bond lengths (Si-Si
and Si-H) in \AA\ and mean bond angles in degrees (Si-Si-Si,
Si-Si-H, Si-H-Si and H-Si-H). The average is taken of 100
configurations over a time span of 0.5 ps. The bridge Si-H-Si
configuration does not exist in BSiH sample while H-Si-H
configuration exist only in ASiH and BSiH samples. The root mean
square deviations in the Si-Si bond length ($\Delta r$) and
Si-Si-Si bond angle distributions ($\Delta \theta$) are also
given.}
\begin{ruledtabular}
\begin{tabular}{c|ccccccccc}
samples&Si-Si&$\Delta r$&Si-H&\vline &Si-Si-Si&$\Delta 
\theta$&Si-Si-H&Si-H-Si&H-Si-H\\
\hline
WWW&2.35&0.08&&&108.57&11.19&&&\\
ASi&2.38&0.13&&&108.23&16.65&&&\\
ASiH&2.39&0.12&1.556&&107.93&17.15&103.20&157.15&107.90\\
BSiH&2.38&0.11&1.535&&108.17&17.37&103.81&-&102.25\\
CSiH&2.36&0.13&1.531&&108.79&15.79&104.67&114.36&-\\
WSiH&2.36&0.11&1.532&&108.80&15.61&102.79&156.75&-\\
\end{tabular}
\end{ruledtabular}
\end{table}

\newpage
\begin{figure}[ht]
\vspace*{-2.5cm} \centerline{\includegraphics{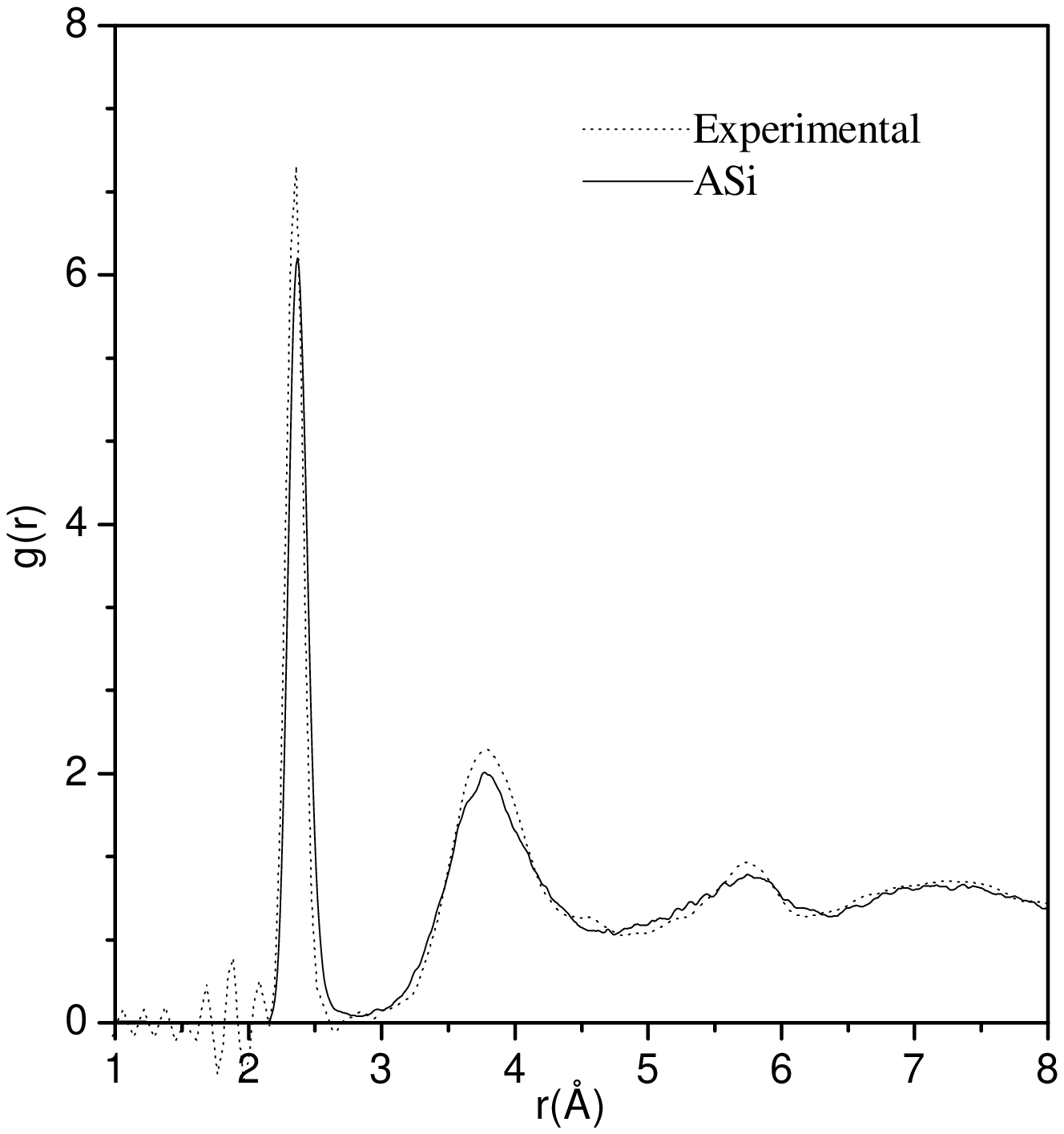}}
\vspace*{-7.4cm} \caption{The Si-Si pair correlation function,
g(r) of pure amorphous silicon sample (ASi) along with the
experimental data \cite{laz99}.}
\end{figure}

\begin{figure}[ht]
\vspace*{-4.0cm} \centerline{\includegraphics{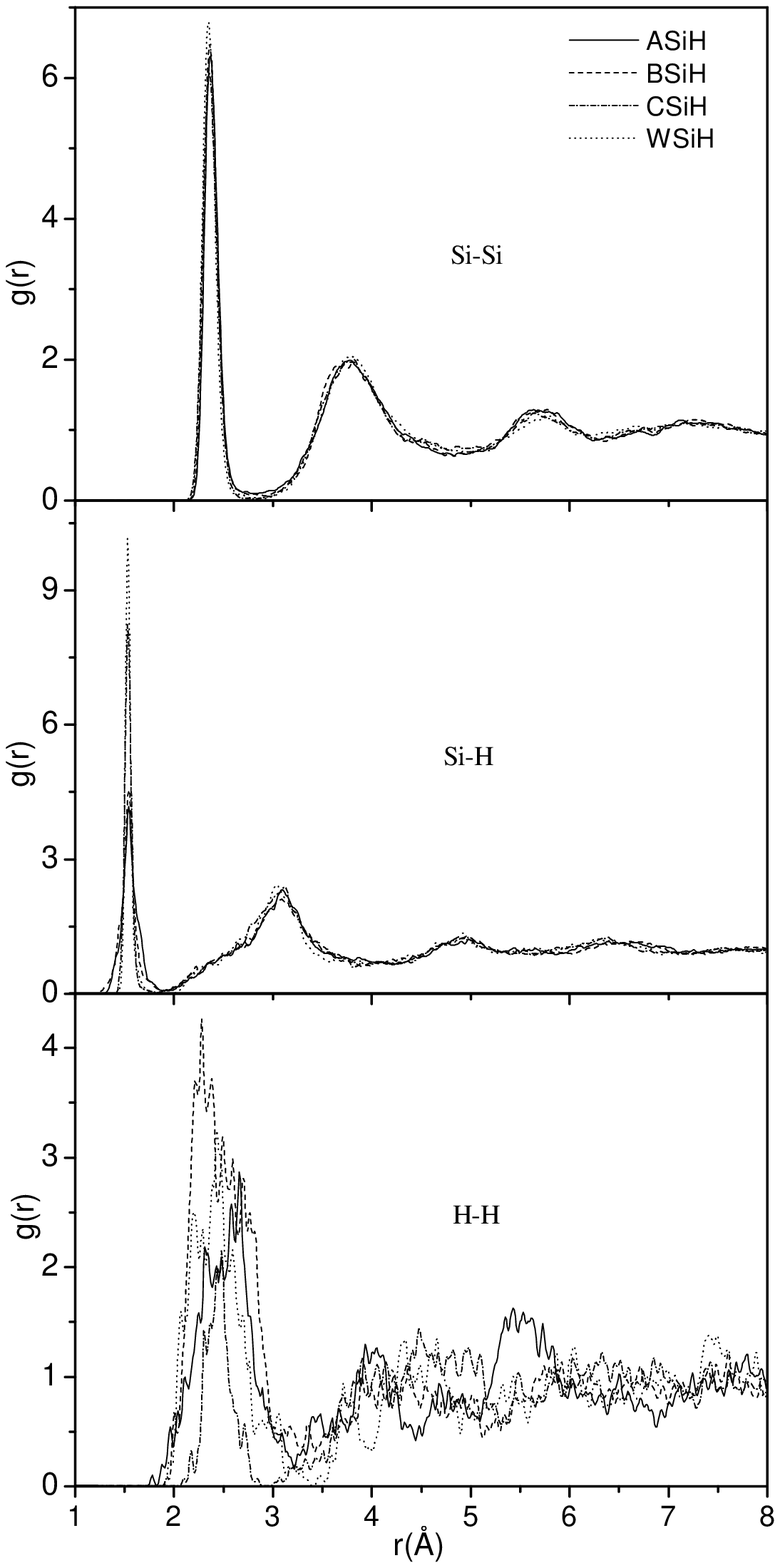}}
\vspace*{-2.2cm} \caption{The partial pair correlation functions,
g(r) of Si-Si, Si-H and H-H correlations in $a$-Si:H samples
(ASiH, BSiH, CSiH and WSiH). The labelling of all the curves in
the middle and bottom panels is same as in the top panel.}
\end{figure}

\begin{figure}[ht]
\vspace*{-3.0cm} \centerline{\includegraphics{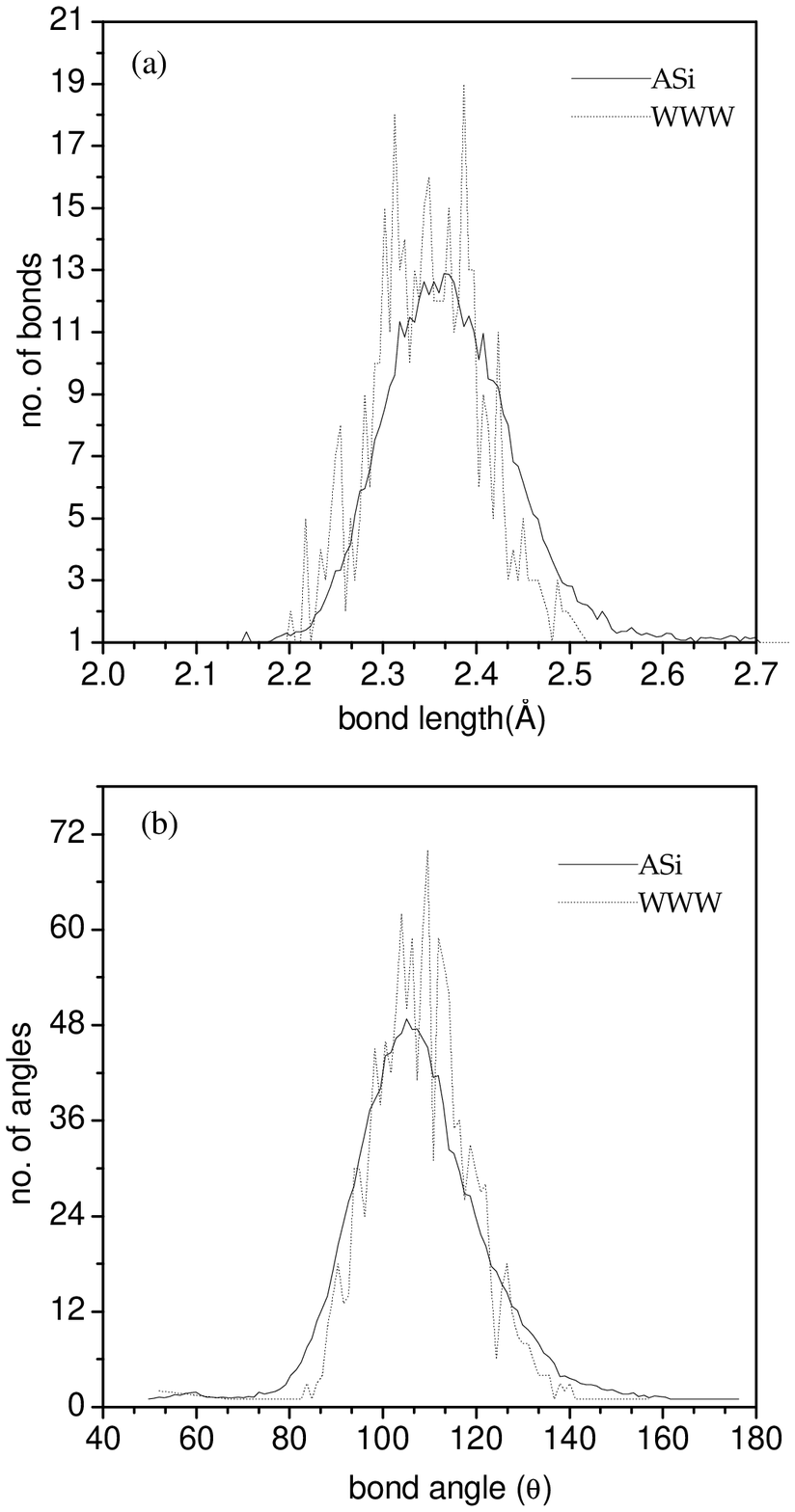}}
\vspace*{-4.0cm} \caption{The distribution functions of ({\bf a})
Si-Si bond length and ({\bf b}) Si-Si-Si bond angle in  ASi and
WWW samples.}
\end{figure}

\begin{figure}[ht]
\vspace*{-2.5cm} \centerline{\includegraphics{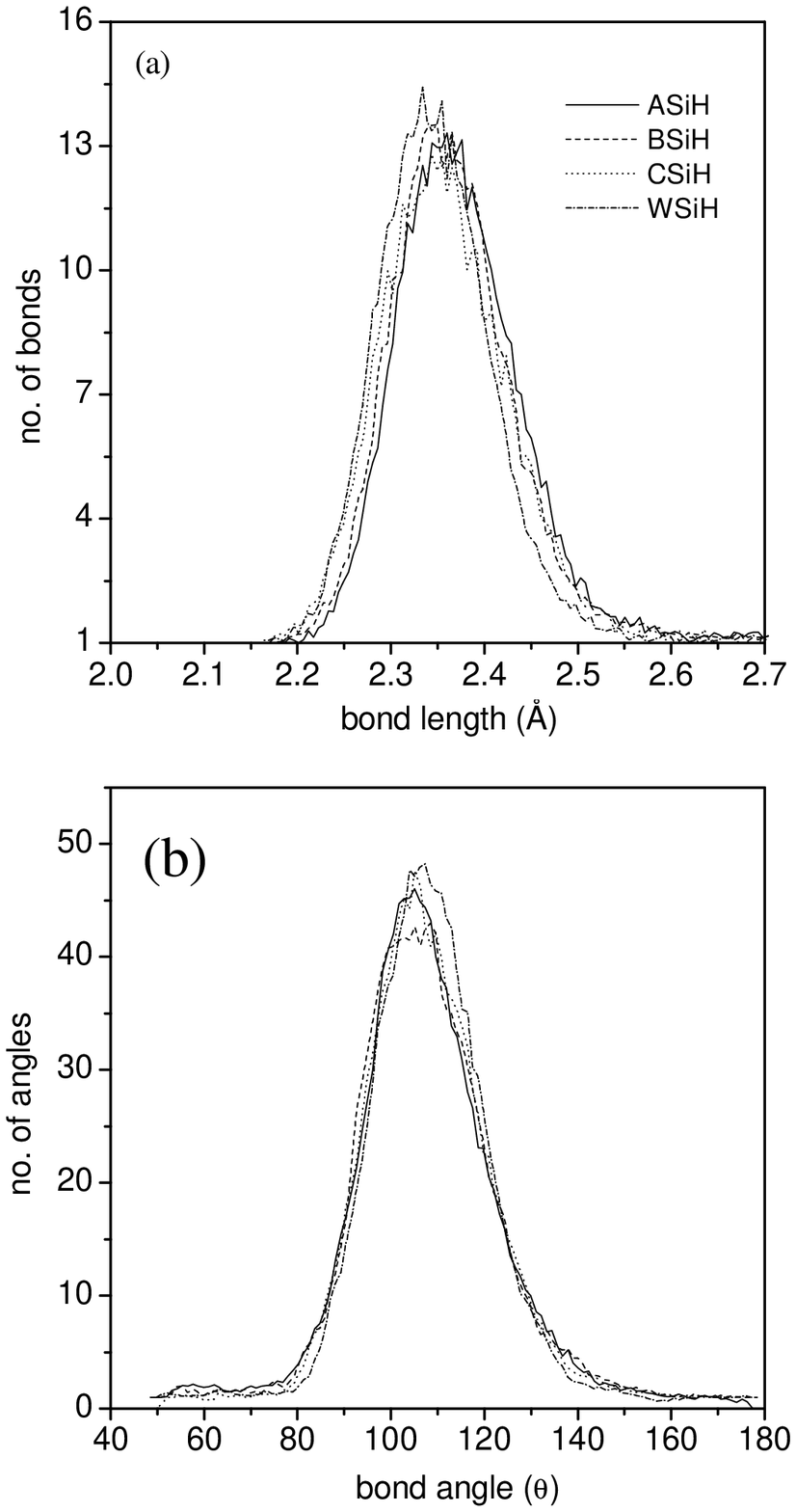}}
\vspace*{-4.0cm} \caption{The distribution functions of ({\bf a})
Si-Si bond length and ({\bf b}) Si-Si-Si bond angle in $a$-Si:H
samples (ASiH, BSiH, CSiH and WSiH). The labelling of all the
curves in the bottom panel is same as in the top panel.}
\end{figure}

\begin{figure}[ht]
\vspace*{-2.5cm} \centerline{\includegraphics{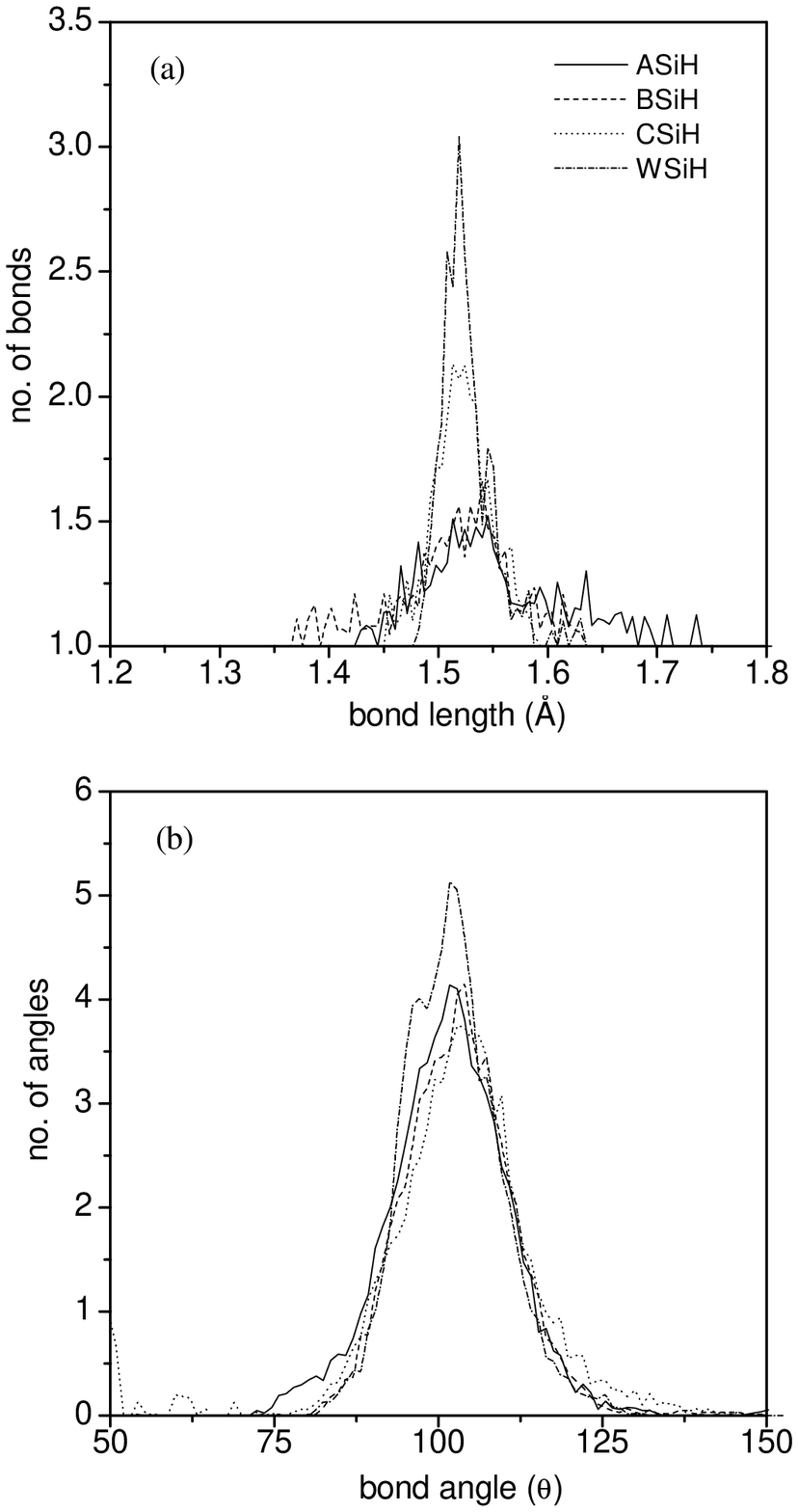}}
\vspace*{-4.2cm} \caption{The distribution functions of ({\bf a})
Si-H bond length, and ({\bf b}) Si-Si-H bond angle in $a$-Si:H
samples (ASiH, BSiH, CSiH and WSiH). The labelling of all the
curves in the bottom panel is same as in the top panel.}
\end{figure}

\begin{figure}[ht]
\vspace*{-3.0cm} \centerline{\includegraphics{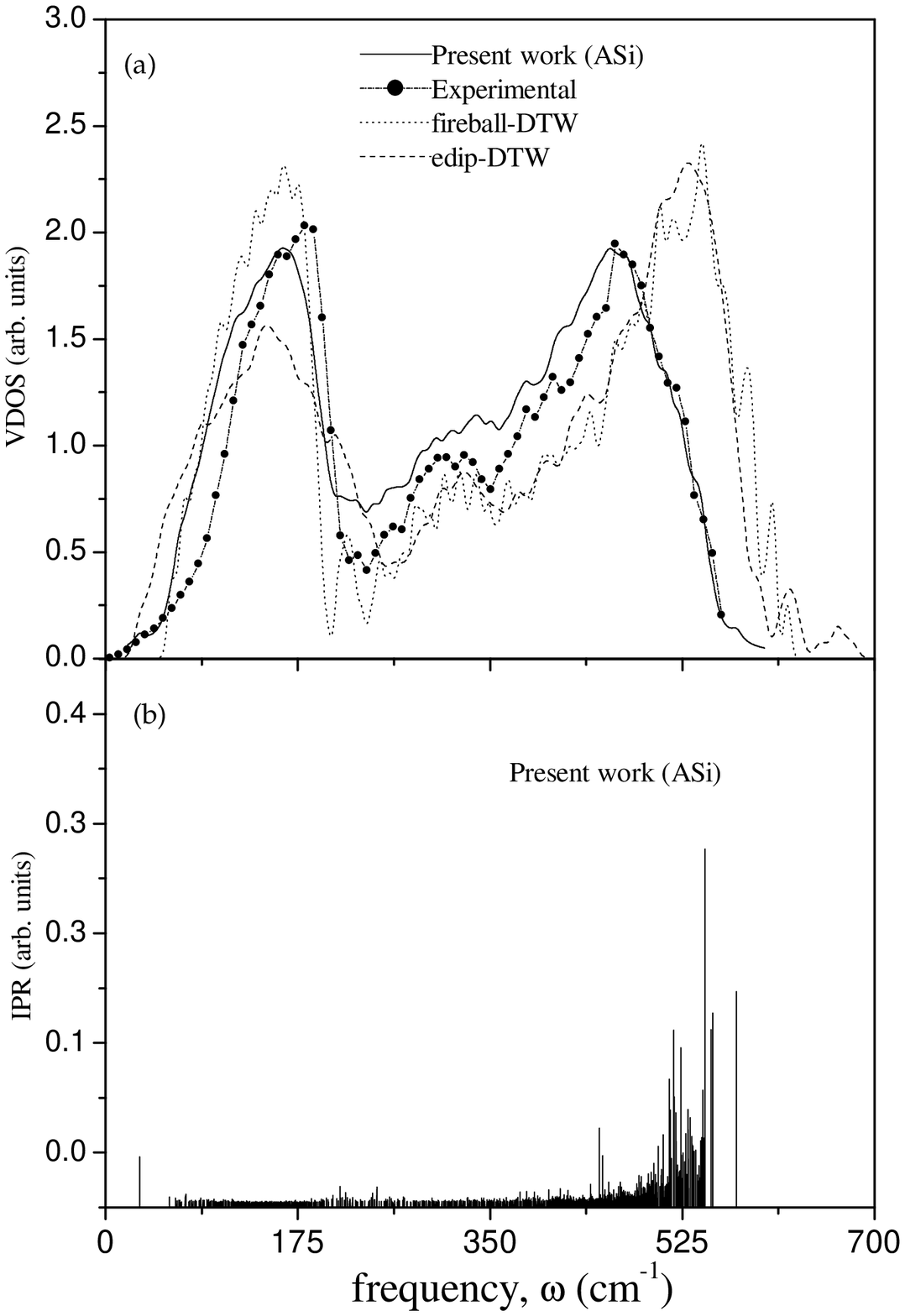}}
\vspace*{-4.2cm} \caption{({\bf a}) The comparison of vibrational
density of states (VDOS) of our sample (ASi) of pure amorphous
silicon with the experimental \cite{kam87} and other model samples
(fireball-DTW \cite{nak98}, edip-DTW \cite{nak00}). ({\bf b}) The
inverse participation ratio (IPR) of ASi sample indicating that
the high frequency modes are localized.}
\end{figure}

\begin{figure}[ht]
\vspace*{-3.50cm} \centerline{\includegraphics{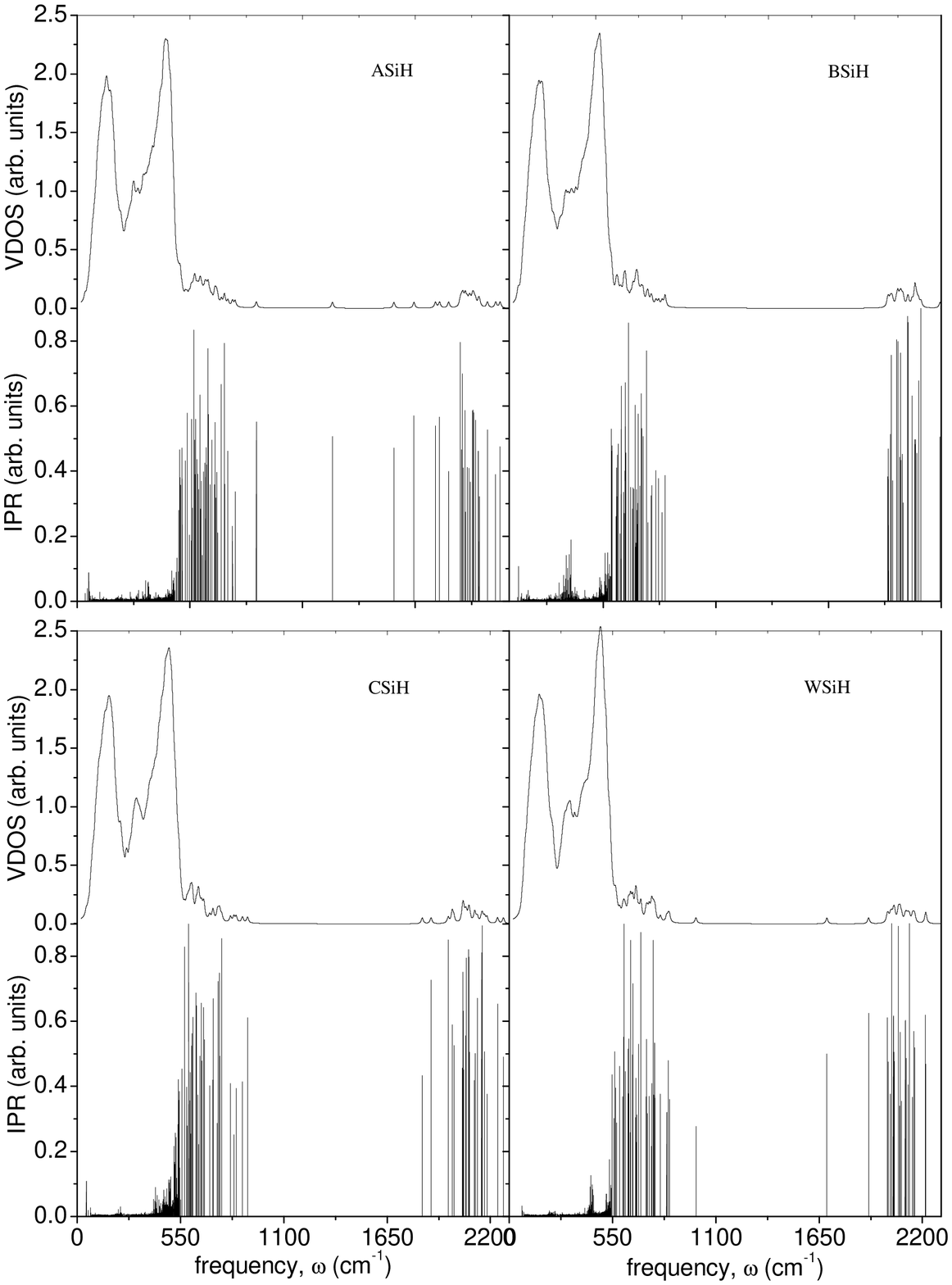}}
\vspace*{-2.0cm} \caption{The VDOS and IPR of vibrational modes in
$a$-Si:H samples (ASiH, BSiH, CSiH and WSiH). For clarity these
are displayed separately for each sample.}
\end{figure}

\begin{figure}[ht]
\vspace*{-4.0cm} \centerline{\includegraphics{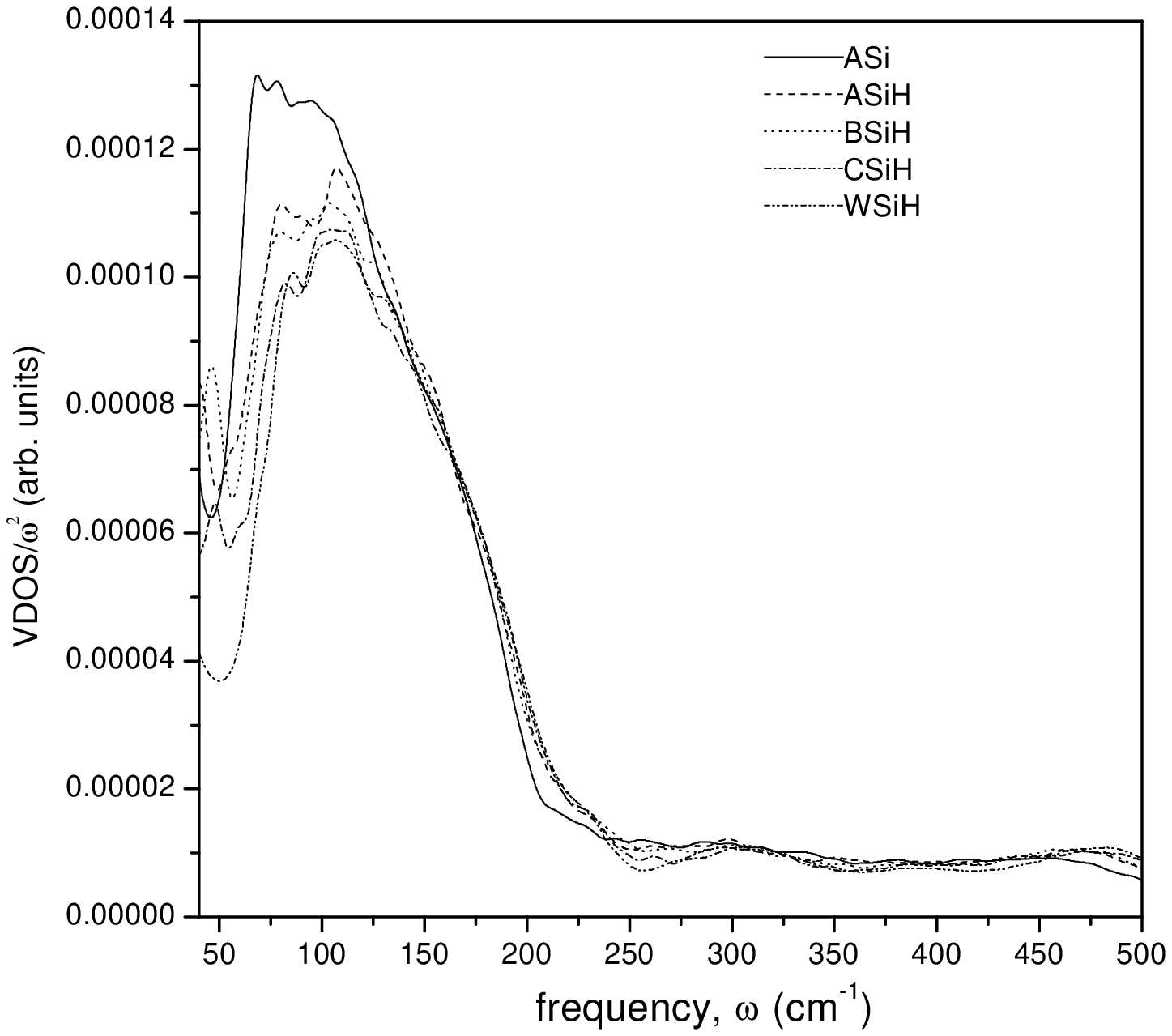}}
\vspace*{-7.5cm} \caption{VDOS/$\omega^{2}$ versus $\omega$ for
model samples of pure and hydrogenated amorphous silicon.}
\end{figure}

\begin{figure}[ht]
\vspace*{-2.7cm} \centerline{\includegraphics{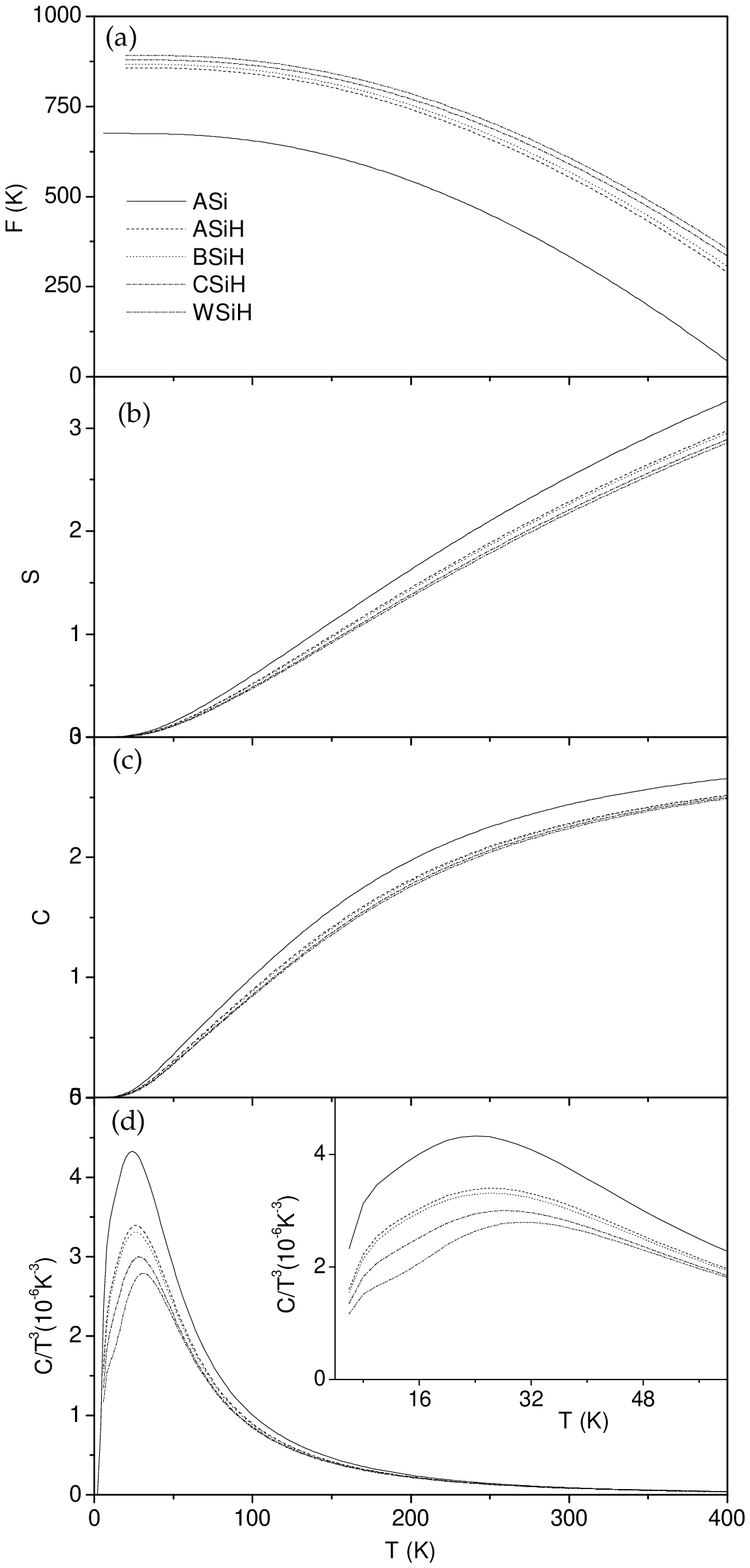}}
\vspace*{-3.3cm} \caption{The temperature (T) dependence of ({\bf
a}) free energy (F), ({\bf b}) vibrational entropy (S) and ({\bf
c}) specific heat (C) calculated from the VDOS within the harmonic
approximation. In ({\bf d}), we have displayed C/T$^{3}$ versus T
curves. The values along y-axis in all panels are per atom in
units of k$_{B}$. The curve descriptions are same in all the
panels as given in ({\bf a}).}
\end{figure}

\begin{figure}[ht]
\vspace*{-4.0cm} \centerline{\includegraphics{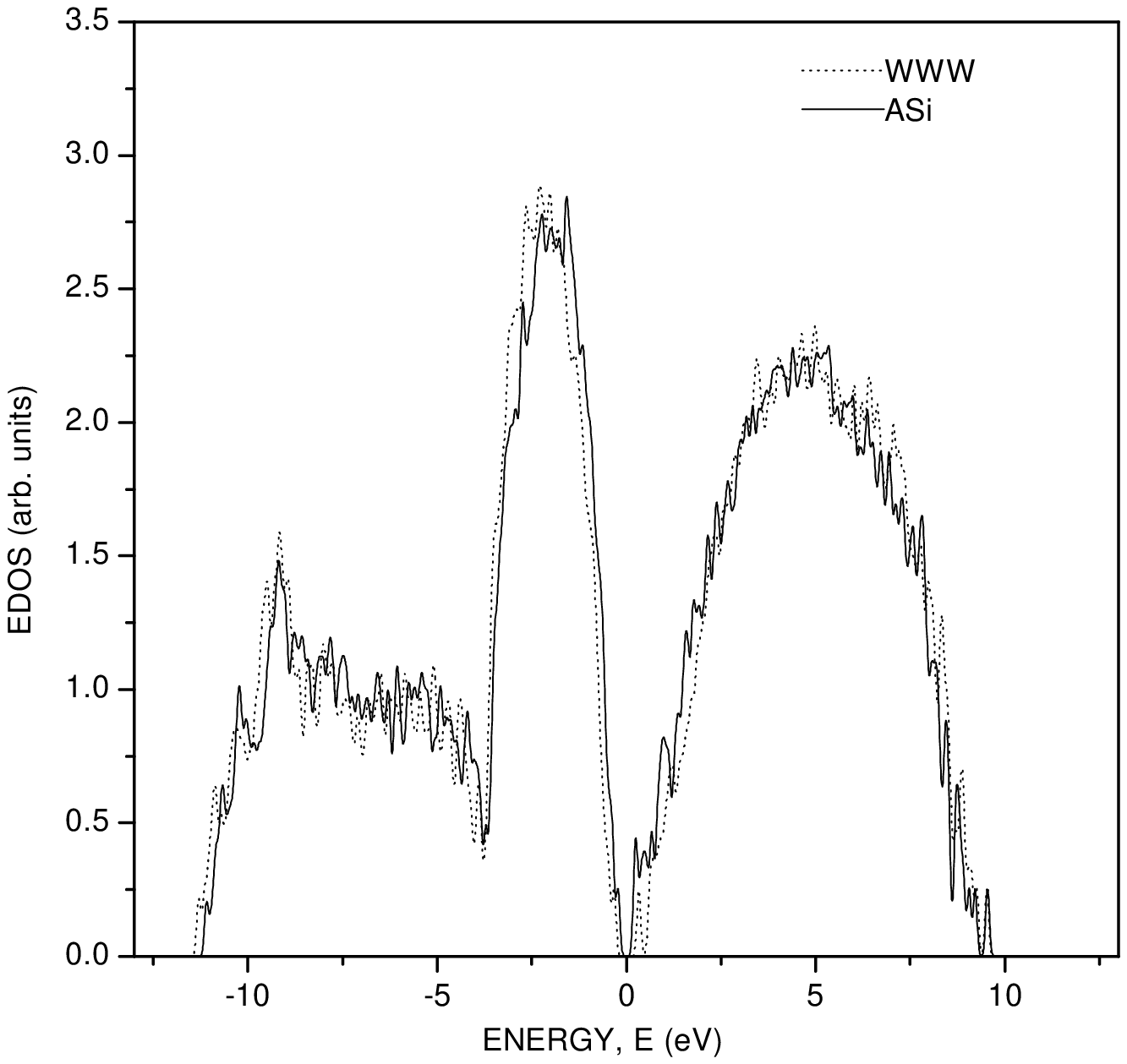}}
\vspace*{-9.0cm} \caption{Electronic density of states (EDOS) for
ASi and WWW samples \cite{wob85} of pure amorphous silicon. E=0 eV
corresponds to the Fermi level.}
\end{figure}

\begin{figure}[ht]
\vspace*{-2.6cm} \centerline{\includegraphics{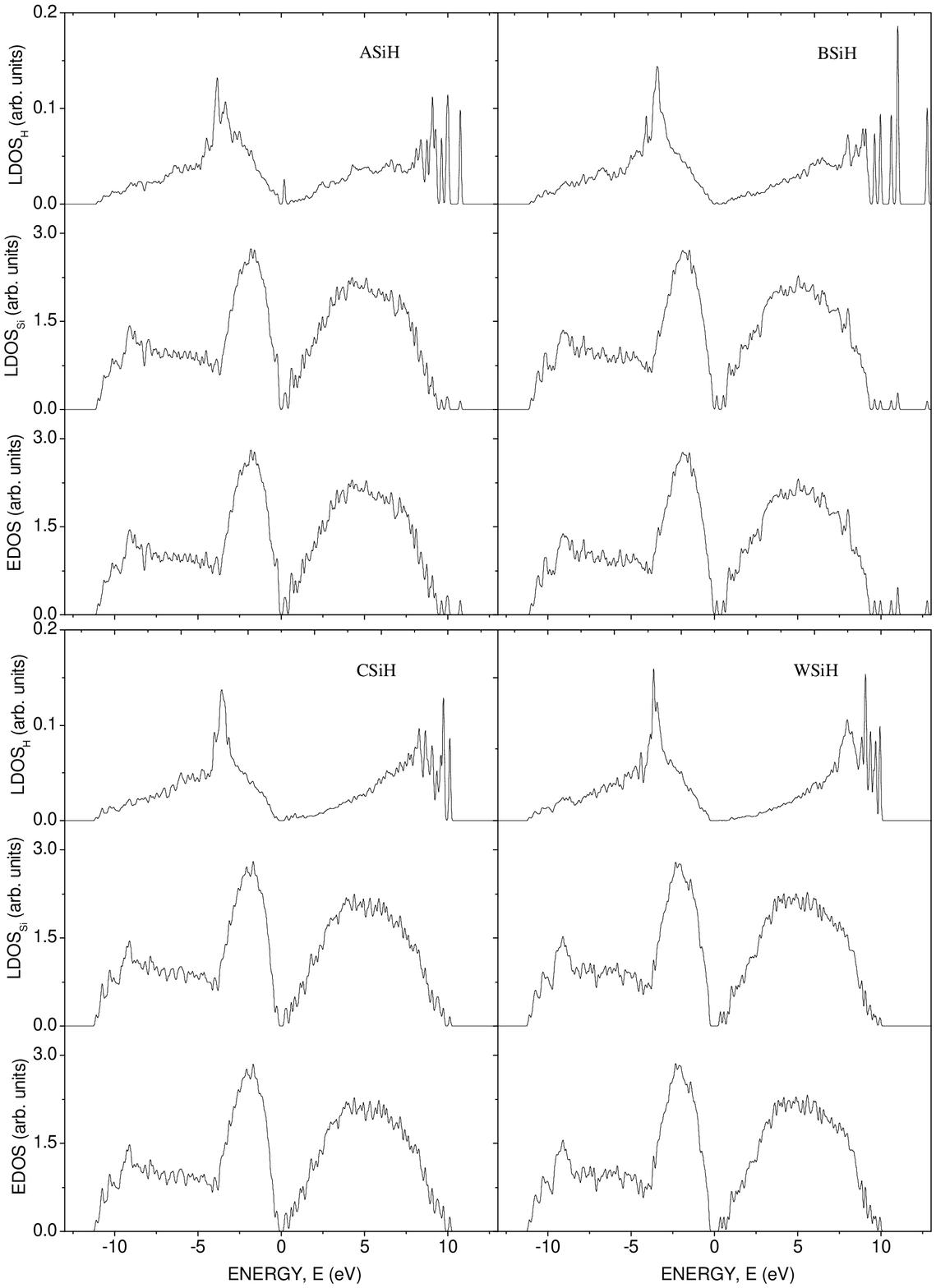}}
\vspace*{-3.0cm} \caption{The local electronic density of states
of hydrogen atoms (LDOS$_{H}$) and silicon atoms (LDOS$_{Si}$) and
total electronic density of states (EDOS) for $a$-Si:H samples
(ASiH, BSiH, CSiH and WSiH). For clarity these are displayed
separately for each sample. E=0 eV corresponds to the Fermi
level.}
\end{figure}

\end{document}